\documentclass[twocolumn]{jpsj2} 

\usepackage{times}

\usepackage{graphicx}

\def\runauthor{Takashi {\sc Hotta}}

\title{%
Microscopic Approach to Magnetism and Superconductivity \\
of $f$-Electron Systems with Filled Skutterudite Structure
}

\author{%
Takashi {\sc Hotta}
}

\inst{%
Advanced Science Research Center,
Japan Atomic Energy Research Institute,
Tokai, Ibaraki 319-1195
}

\recdate{\today}

\abst{%
In order to gain a deep insight into $f$-electron properties of
filled skutterudite compounds from a microscopic viewpoint,
we investigate the multiorbital Anderson model
including Coulomb interactions, spin-orbit coupling,
and crystalline electric field effect.
First we examine the local $f$-electron state in detail
in comparison with the results of $LS$ and $j$-$j$ coupling schemes.
For each case of $n$=1$\sim$13, where $n$ is the number of
$f$ electrons per rare-earth ion,
the model is analyzed by using the numerical
renormalization group (NRG) method to evaluate
magnetic susceptibility and entropy of $f$ electron.
In particular, for the $f^2$-electron system corresponding to
the Pr-based filled skutterudite,
it is found that magnetic fluctuations significantly remain
at low temperatures, even when the ground state is $\Gamma_1$ singlet,
if $\Gamma_4^{(2)}$ triplet is the excited state
with small excitation energy.
In order to make further step to construct a simplified model
which can be treated even in a periodic system,
we also analyze the Anderson model constructed based on the $j$-$j$
coupling scheme by using the NRG method.
It is clearly observed that the magnetic properties
are quite similar to those of the original Anderson model.
Then, we construct an orbital degenerate Hubbard model based on
the $j$-$j$ coupling scheme to investigate the mechanism of
superconductivity of filled skutterudites.
In the 2-site model, we carefully evaluate the superconducting
pair susceptibility for the case of $n$=2 and find that
the susceptibility for off-site Cooper pair is
clearly enhanced only in a transition region
in which the singlet and triplet ground states are interchanged.
We envision a scenario that unconventional superconductivity
induced by magnetic fluctuations may occur in the $f^2$-electron
system with $\Gamma_1$ ground state such as Pr-based
filled skutterudite compounds.
}

\kword{%
Filled skutterudites, Multiorbital Anderson model,
Numerical renormalization group method,
Magnetic fluctuations, Orbital fluctuations,
Unconventional superconductivity}

\begin{document}
\maketitle

%
%
\section{Introduction}

Recently $f$-electron compounds with filled skutterudite structure,
expressed as RT$_4$X$_{12}$ (R=rare-earth ion, T=transition metal ion,
and X=pnictogen), have attracted much attention in the research field
of condensed matter physics.\cite{Sato}
Since the filled skutterudite compounds exhibit huge thermopower,
they have been focused as thermoelectric materials,
mainly for the purpose of industrial application.
However, after the discovery of heavy fermion behavior
and superconductivity in PrOs$_4$Sb$_{12}$,\cite{Maple,Bauer}
the filled skutterudite compound has been an important target of
active investigations also from the viewpoint of basic material
science, since this is the first Pr-based $f^2$-electron
system that shows heavy-fermion superconductivity.

From experimental results on specific heat,\cite{Aoki1}
magnetization,\cite{Tayama} and neutron scattering,\cite{Kohgi,Kuwahara}
it has been recently confirmed that the ground state of
PrOs$_4$Sb$_{12}$ is $\Gamma_1$ singlet,
while the first excited state is $\Gamma_4^{(2)}$ triplet
with very small excitation energy less than 10K.
Note that the filled skutterudite has cubic symmetry
described by $T_{\rm h}$, not $O_{\rm h}$ point group.\cite{Takegahara}
In such a material with non-magnetic $\Gamma_1$ singlet ground state,
conventional superconductivity is naively expected.
However, the superconductivity of PrOs$_4$Sb$_{12}$ is suggested to
be unconventional from NMR experiment,\cite{Kotegawa}
since there is no coherence peak in $1/T_1$,
where $T_1$ is nuclear spin-lattice relaxation time,
just below the superconducting transition temperature $T_{\rm c}$.

Here one may expect that PrOs$_4$Sb$_{12}$ belongs to
the group of unconventional $d$-wave superconductors,
as has been frequently observed in strongly correlated electron systems
including Ce- and U-based heavy fermion superconductors.\cite{Yanase}
However, $1/T_1$ of PrOs$_4$Sb$_{12}$ does $not$ obey the so-called
$T^3$ behavior,\cite{Kotegawa} characteristic of the line-node gap
function in the $d$-wave pairing.
Rather, the isotropic superconducting gap is consistent with
the experimental results, in which $1/T_1T$ exponentially decreases
even crossing $T_{\rm c}$.
In addition, the superconductivity of filled skutterudite compounds
exhibit several exotic features, for instance,
point-node behavior in the gap function,\cite{Izawa,Chia}
multiple superconducting phases,\cite{Izawa}
and the breaking of the time-reversal symmetry
detected by $\mu$SR experiment.\cite{Aoki2}
Then, it has been highly requested to elucidate theoretically
the mechanism of such exotic superconductivity of PrOs$_4$Sb$_{12}$.

Let us turn our attention to Pr-based filled skutterudites
including other transition metal ions and pnictogens.
It has been found that the ground-state properties are changed
sensitively depending on T and X in PrT$_4$X$_{12}$.
As mentioned above, PrOs$_4$Sb$_{12}$ exhibits unconventional
superconductivity,
while PrRu$_4$Sb$_{12}$ is confirmed to be conventional $s$-wave
superconductor from Sb-NQR experiment.\cite{Yogi}
Interestingly enough, due to the substitution of transition metal ion,
superconducting properties in ${\rm Pr(Os_{1-x}Ru_{x})_4Sb_{12}}$
are changed from unconventional to conventional ones.\cite{Frederick}
When pnictogens are substituted,
the ground-state nature is also drastically changed.
For instance, PrOs$_4$P$_{12}$ is non-magnetic metal,\cite{Sekine}
PrRu$_4$P$_{12}$ exhibits metal-insulator transition,
\cite{Sekine,Harima1}
and PrFe$_4$P$_{12}$ shows exotic quadrupolar ordering.\cite{Aoki3}
It is also an important task imposed on theoretical investigations
to clarify what is a key issue to control the electronic properties
of Pr-based filled skutterudites.

In addition to Pr-based compounds, numerous kinds of filled skutterudite
materials have been synthesized by expert techniques.
Those compounds actually exhibit varieties of electronic properties:
La-based filled skutterudite materials are known to be conventional
BCS superconductors.\cite{Kotegawa}
Ce-based filled skutterudites are Kondo semi-conductors
with the gap up to thousand Kelvins.
For the skutterudites including rare-earth ions other than
La, Ce, and Pr, the ferromagnetic ground state has been
frequently observed, for instance, in
RFe$_4$P$_{12}$ for R=Nd, Sm, Eu, Gd, Tb, Dy, and Ho.
However, in some Sm-, Gd-, and Tb-based skutterudites such as
SmOs$_4$Sb$_{12}$\cite{Giri}, GdRu$_4$P$_{12}$,\cite{Sekine3}
and TbRu$_4$P$_{12}$,\cite{Sekine3}
antiferromagnetic ground states have been also found.
It is another characteristic issue of filled skutterudites
that the magnetic properties are changed depending on the
number of $f$-electrons $n$ per rare-earth ion.

Even in the above brief survey of the properties of filled skutterudite
compounds, we can easily understand that this group of materials
exhibits richness, as revealed by experimental effort.
However, theoretical investigations on magnetism and
superconductivity for $f^n$-electron systems with $n$$>$1 have been
almost limited to the phenomenological level
based on the Ginzburg-Landau theory, mainly due to complexity
in the many-body problem originating from the competition
among strong spin-orbit coupling, Coulomb interactions,
and the effect of crystalline electric field (CEF).
For the purpose to include such interactions,
an $LS$ coupling scheme has been widely used,
but it is not possible to apply standard quantum-field theoretical
technique in the $LS$ coupling scheme,
since Wick's theorem does not hold.

In order to overcome such a difficulty, it has been proposed to
construct a microscopic model for $f$-electron systems
from a $j$-$j$ coupling scheme.\cite{Hotta1}
Since individual $f$-electron states are clearly defined
in the $j$-$j$ coupling scheme, it is convenient for the inclusion
of many-body effects using standard theoretical techniques.
In fact, based on such a microscopic model, the following points
have been clarified:
Key role of orbital degree of freedom for superconductivity
of CeTIn$_5$ and PuTGa$_5$ with transition metal ion T,
\cite{Hotta1,Maehira,Takimoto}
an odd-parity triplet pair induced by Hund's rule interaction,
\cite{Hotta2}
the complex magnetic structure of UTGa$_5$\cite{Hotta3}
and NpTGa$_5$\cite{Onishi} based on a spin-orbital model,
and microscopic origin of octupole ordering of NpO$_2$.\cite{Kubo}

When we attempt to apply the microscopic model based on the $j$-$j$
coupling scheme to the filled skutterudite compound,
unfortunately, there are several problems which should be clarified.
First of all, the application of the $j$-$j$ coupling scheme to
the rare-earth materials seems questionable except for cerium compounds,
since the Coulomb interactions are larger than the spin-orbit coupling
in $4f$-electron systems.
Second, it is still unclear how to include the CEF effect
for $f^n$-electron systems in the $j$-$j$ coupling scheme,
in which the CEF effect is just a one-electron potential
determined for the $f^1$-electron system.\cite{Hotta1}
It is necessary to show explicitly how the $j$-$j$ coupling scheme
reproduces the electronic properties of $f^n$-electron systems.

In this paper, first we analyze in detail the $f$-electron term
including spin-orbit coupling, Coulomb interactions, and CEF effect.
In particular, we carefully compare the local $f$-electron states
with those of the $LS$ and $j$-$j$ coupling schemes.
We also show the CEF energy levels for $n$=1$\sim$13
under the CEF potential of $T_{\rm h}$ point group.
By further adding the conduction electron and hybridization terms,
we obtain the Anderson model and analyze it numerically.
Then, the Curie-law behavior in magnetic susceptibility is found
even in the $f^2$-electron system, when $\Gamma_1$ singlet is the
local ground state and $\Gamma_4^{(2)}$ triplet is
the excited state with small excitation energy.
The same tendency is observed in the Anderson model constructed
from the $j$-$j$ coupling scheme. Then, we further construct
the Hubbard-like model based on the $j$-$j$ coupling scheme
and evaluate the superconducting pair susceptibility.
The numerical results on the model suggest that
anisotropic Cooper-pair may appear in a region in which $\Gamma_1$
singlet and $\Gamma_4^{(2)}$ triplet states are interchanged.

The organization of this paper is as follows.
In Sec.~2, we discuss in detail the properties of local $f$-electron term
in comparison with those of the $LS$ and $j$-$j$ coupling schemes.
Then, we show the energy levels by changing the $f$-electron number $n$
between $n$=1 and 13 for the fixed CEF parameters of the $T_{\rm h}$
point group.
In Sec.~3, the Anderson model is introduced for filled skutterudites
and it is analyzed by using the numerical renormalization group
(NRG) method. The results for magnetic susceptibility and entropy are
shown for $n$=1$\sim$13.
In order to reduce effectively the number of $f$-orbitals,
we also introduce the Anderson model in the $j$-$j$ coupling scheme by
focusing on the case of $n$=2, i.e., the case of Pr-based
filled skutterudites.
The model is analyzed by using the NRG method and it is shown that
the essential point of the results in the original Anderson model can be
correctly captured even in the model based on the $j$-$j$ coupling scheme.
In Sec.~4, we construct an orbital degenerate Hubbard model appropriate
for filled skutterudites.
By evaluating carefully the superconducting pair susceptibility of the
model, we clearly show that the pair susceptibility is enhanced only
in a narrow region in which the ground state is changed between singlet
and triplet states.
In Sec.~5, we will summarize the paper.

%
%
\section{Local f-electron state}

In this section, let us examine in detail the properties of
local $f$-electron states in comparison with those obtained
in the $LS$ and $j$-$j$ coupling schemes.

\subsection{Local $f$-electron Hamiltonian}

In general, the local $f$-electron term is composed of three parts as
\begin{eqnarray}
 H_{\rm f} = H_{\rm C} + H_{\rm so} + H_{\rm CEF},
\end{eqnarray}
where $H_{\rm C}$ is the Coulomb interaction term, written as
\begin{eqnarray}
 H_{\rm C} &=& \sum_{m_1 \sim m_4}\sum_{\sigma_1, \sigma_2}
 I_{m_1,m_2,m_3,m_4} \nonumber \\
 &\times& f_{m_1\sigma_1}^{\dag}f_{m_2\sigma_2}^{\dag}
 f_{m_3\sigma_2}f_{m_4\sigma_1}.
\end{eqnarray}
Here $f_{m\sigma}$ is the annihilation operator for $f$-electron
with spin $\sigma$ and angular momentum $m$(=$-3$,$\cdots$,3)
and $\sigma$=+1 ($-$1) for up (down) spin.
The Coulomb integral $I_{m_1,m_2,m_3,m_4}$ is known to be written
in the form of
\begin{eqnarray}
  I_{m_1,m_2,m_3,m_4}=\sum_{k=0}^{6} F^k c_k(m_1,m_4)c_k(m_2,m_3),
\end{eqnarray}
where the sum on $k$ includes only even values ($k$=0, 2, 4, and 6),
$F^k$ is the Slater-Condon parameter
including the complex integral of the radial function,\cite{Slater1,Condon}
and $c_k$ is the Gaunt coefficient,\cite{Gaunt,Racah2}
which is tabulated in the standard textbooks of quantum
mechanics.\cite{Slater2}
It is convenient to express the Slater-Condon parameters as
\begin{eqnarray}
 \begin{array}{l}
    F^0=A+15C+9D/7, \\
    F^2=225(B-6C/7+D/42), \\
    F^4=1089(5C/7+D/77), \\
    F^6=(429/5)^2\cdot(D/462),
 \end{array}
\end{eqnarray}
where $A$, $B$, $C$, and $D$ are
the Racah parameters.\cite{Racah2}

The spin-orbit coupling term, $H_{\rm so}$, is given by
\begin{eqnarray}
  H_{\rm so} = \lambda \sum_{m,\sigma,m',\sigma'}
  \zeta_{m,\sigma,m',\sigma'} f_{m\sigma}^{\dag}f_{m'\sigma'},
\end{eqnarray}
where $\lambda$ is the spin-orbit interaction
and the matrix elements are expressed by
\begin{eqnarray}
 \begin{array}{l}
 \zeta_{m,\sigma,m,\sigma}=m\sigma/2,\\
 \zeta_{m+1,\downarrow,m,\uparrow}=\sqrt{12-m(m+1)}/2,\\
 \zeta_{m-1,\uparrow,m,\downarrow}=\sqrt{12-m(m-1)}/2,
 \end{array}
\end{eqnarray}
and zero for other cases.

The CEF term $H_{\rm CEF}$ is given by
\begin{eqnarray}
  \label{Eq:CEF}
  H_{\rm CEF} = \sum_{m,m',\sigma} B_{m,m'}
  f_{m\sigma}^{\dag}f_{m'\sigma},
\end{eqnarray}
where $B_{m,m'}$ is determined from the table of Hutchings
for angular momentum $\ell$=3,\cite{Hutchings}
since we are now considering the potential for $f$ electron.
For filled skutterudites with $T_{\rm h}$ symmetry,\cite{Takegahara}
$B_{m,m'}$ is expressed by using three CEF parameters
$B_{40}$, $B_{60}$, and $B_{62}$ as
\begin{eqnarray}
  \begin{array}{l}
    B_{3,3}=B_{-3,-3}=180B_{40}+180B_{60}, \\
    B_{2,2}=B_{-2,-2}=-420B_{40}-1080B_{60}, \\
    B_{1,1}=B_{-1,-1}=60B_{40}+2700B_{60}, \\
    B_{0,0}=360B_{40}-3600B_{60}, \\
    B_{3,-1}=B_{-3,1}=60\sqrt{15}(B_{40}-21B_{60}),\\
    B_{2,-2}=300B_{40}+7560B_{60},\\
    B_{3,1}=B_{-3,-1}=24\sqrt{15}B_{62},\\
    B_{2,0}=B_{-2,0}=-48\sqrt{15}B_{62},\\
    B_{1,-1}=-B_{3,-3}=360B_{62}.
  \end{array}
\end{eqnarray}
Note the relation of $B_{m,m'}$=$B_{m',m}$.
Following the traditional notation, we define
\begin{eqnarray}
  \begin{array}{l}
    B_{40}=Wx/F(4),\\
    B_{60}=W(1-|x|)/F(6),\\
    B_{62}=Wy/F^t(6),
  \end{array}
\end{eqnarray}
where $x$ and $y$ specify the CEF scheme for $T_{\rm h}$ point group,
while $W$ determines an energy scale for the CEF potential. Although
$F(4)$, $F(6)$, and $F^t(6)$ have not been determined
uniquely,\cite{Takegahara2} in Eq.~(\ref{Eq:CEF}) we choose
$F(4)$=15, $F(6)$=180, and $F^t(6)$=24 for $\ell$=3.\cite{Hutchings}

We note that the CEF potential is originally given
by the sum of electrostatic energy from the ligand ions
at the position of $f$-electron ion,
leading to the one-electron potential acting on the charge distribution
of $f$-orbitals, as expressed by Eq.~(\ref{Eq:CEF}).
Thus, in principle, it is $not$ necessary to change the CEF potential
depending on the $f$-electron number.
As we will see in the following subsections, the CEF schemes
for $n$=1$\sim$13 are automatically reproduced by diagonalizing
the local $f$-electron term $H_{\rm f}$,
once we fix the CEF parameters in the form of one-electron
potential Eq.~(\ref{Eq:CEF}).

\subsection{Comparison with $LS$ and $j$-$j$ coupling schemes}

Before proceeding to the discussion on the energy levels of
$T_{\rm h}$ point group for $n$=1$\sim$13,
we compare the electronic states of $H_{\rm f}$
with those of $LS$ and $j$-$j$ coupling schemes,
since it is quite instructive to understand the meanings
of the CEF potential in $f$-electron systems.
We introduce ``$U$'' as an energy scale
for the Racah parameters, $A$, $B$, $C$, and $D$.
In this section, $U$ is the energy unit, which is typically
considered to be 1 eV.

In general, in $f$-electron systems, the magnitude of
the CEF potential is smaller than both spin-orbit coupling and
Coulomb interactions.
Thus, it is reasonable to consider that $W$ is always
much smaller than $\lambda$ and $U$.
However, there occur two situations, depending on the order
to take the limits of $\lambda/W$$\rightarrow$$\infty$ and
$U/W$$\rightarrow$$\infty$.
When the limit of $U/W$$\rightarrow$$\infty$ is first taken and
then we consider the effect of the spin-orbital coupling
$\lambda$, we arrive at the $LS$ coupling scheme.
On the other hand, it is also possible to take first
the infinite limit of $\lambda/W$. After that,
we include the effect of Coulomb interaction,
leading to the $j$-$j$ coupling scheme.
In the present local $f$-electron term $H_{\rm f}$,
it is easy to consider two typical situations for $f$-electron problems,
$|W|$$\ll$$\lambda$$<$$U$ and $|W|$$\ll$$U$$<$$\lambda$, corresponding to
the $LS$ and $j$-$j$ coupling schemes, respectively.

\begin{figure}[t]
\begin{center}
\includegraphics[width=7truecm]{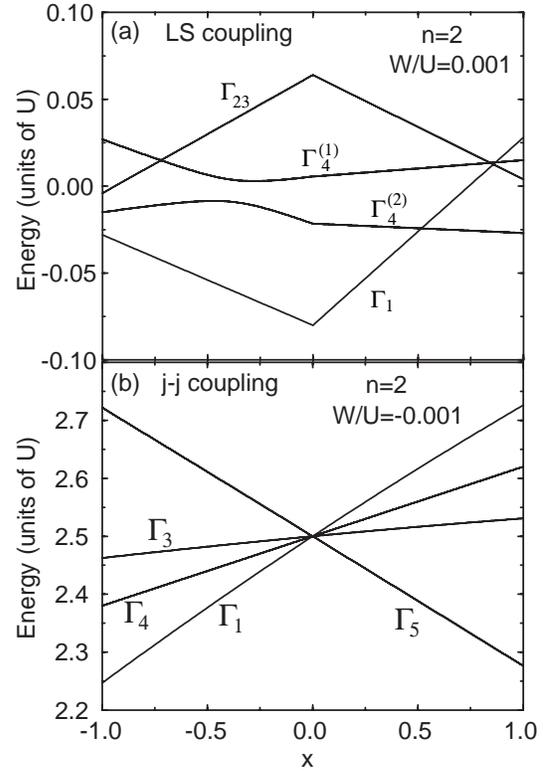}
\caption{Energies of $f$ electrons as functions of $x$ for
(a) the $LS$ coupling and (b) the $j$-$j$ coupling schemes for $n$=2.
The magnitude of the CEF potential energy is fixed as $|W|/U$=0.001.}
\end{center}
\end{figure}

Let us consider here the case of $n$=2 as a typical example of
the comparison with the two schemes.
In the $LS$ coupling scheme for the $f^2$-electron system,
we obtain the ground-state level as $^{3}H$ with $S$=1 and $L$=5
from the Hund's rules, where $S$ and $L$ denote sums of each
$f$-electron spin and angular momentum, respectively.
Upon further including the spin-orbit interaction, the ground state
is specified by $J$=4 expressed as $^3H_4$ in the traditional notation.
Note that the total angular momentum $J$ is given by $J$=$|L-S|$ and
$J$=$L+S$ for $n$$<$7 and $n$$>$7, respectively.
In order to consider further the CEF effect, by consulting with
the table of Hutchings for the case of $J$=4, we easily obtain
the nine eigen values, including $\Gamma_1$ singlet, $\Gamma_{23}$
doublet, and two kinds of $\Gamma_4$ triplets, as shown in Fig.~1(a).
We note that in the $LS$ coupling scheme,
we set $F(4)$=60, $F(6)$=1260, $F^t(6)$=30, $y$=0.05, and $W/U$=0.001.
Note also that two $\Gamma_4$ triplets in $T_{\rm h}$ are obtained
by the mixtures of $\Gamma_4$ and $\Gamma_5$ triplets
in the $O_{\rm h}$ point group.

In the $j$-$j$ coupling scheme, on the other hand, first we take
the infinite limit of $\lambda$.
Thus, we consider only the $j$=5/2 sextet, where $j$ denotes the total
angular momentum of one $f$-electron.
In the $f^2$-electron system, two electrons are accommodated in
the sextet, leading to fifteen eigen states including
$J$=4 nontet, $J$=2 quintet, and $J$=0 singlet.
Due to the effect of Hund's rule coupling, $J$=4 nontet becomes
the ground state.
When we further include the CEF potential, it is necessary to reconsider
the accommodations of two electrons in the $f^1$-electron potential
with $\Gamma_7$ doublet and $\Gamma_8$ quartet.
Note that the difference between $O_{\rm h}$ and $T_{\rm h}$ does not
appear for the case of $J$=5/2, namely in the $j$-$j$ coupling scheme.
Thus, in the $j$-$j$ coupling schemes, except for the energy scale $W$,
only relevant CEF parameter is $x$, leading to the level splitting
between $\Gamma_7$ doublet and $\Gamma_8$ quartet.
For the $j$-$j$ coupling scheme,
we set $F(4)$=60 and $W/U$=$-0.001$.
Note that the minus sign in $W$ is added for the purpose of
easy comparison with the $LS$ coupling scheme.
As shown in Fig.~1(b), the $J$=4 nontet is split into
$\Gamma_1$ singlet, $\Gamma_{3}$ doublet,
$\Gamma_4$ triplet, and $\Gamma_5$ triplet.
For simplicity, we use the notations for irreducible representations
of $O_{\rm h}$ symmetry.
The ground state for $x$$>$0 is $\Gamma_5$ triplet composed of
a couple of $\Gamma_8$ electrons, while for $x$$<$0, it is $\Gamma_1$
singlet which is mainly composed of two $\Gamma_7$ electrons.
Note that for $x$$>$0, the first excited state is $\Gamma_4$ triplet,
composed of $\Gamma_7$ and $\Gamma_8$ electrons.

\begin{figure}[t]
\begin{center}
\includegraphics[width=7truecm]{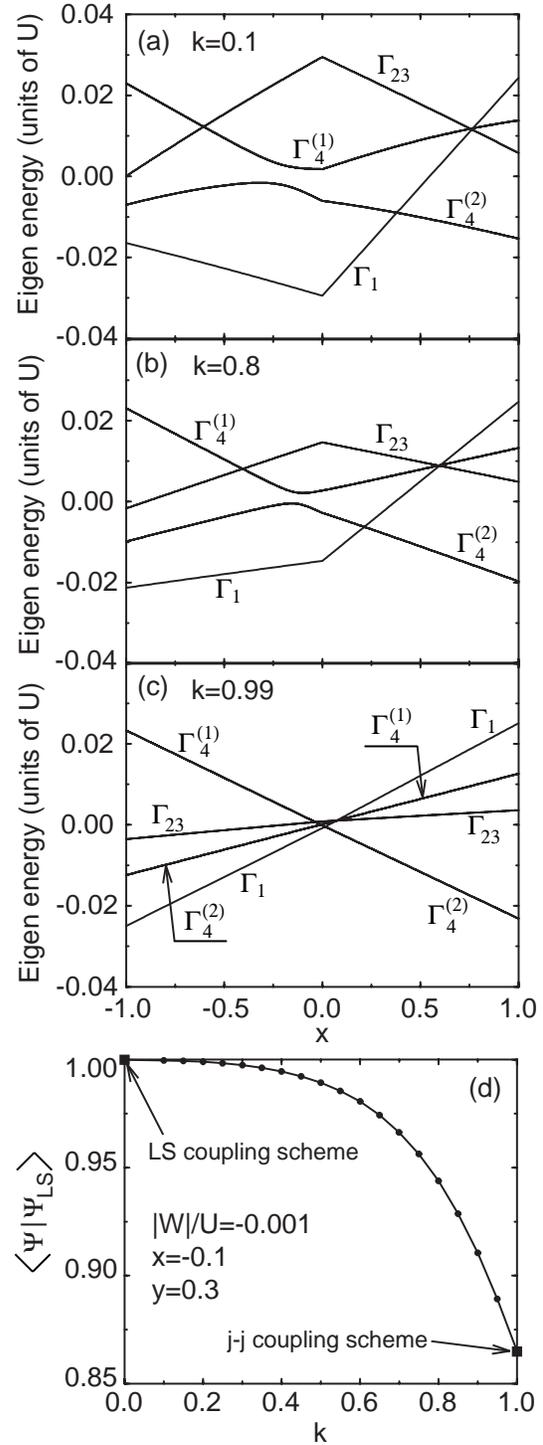}
\caption{Eigen energies of $H_{\rm f}$ as functions of $x$ for
(a) $k$=0.1, (b) $k$=0.8, and (c) $k$=0.99.
Racah parameters are set as $A/U$=10, $B/U$=0.3,
$C/U$=0.1, and $D/U$=0.05.
CEF potentials are given by $W/U$=$-0.001$ and $y$=0.3.
Solid squares at $k$=0 and 1 are obtained separately from
the $LS$ and $j$-$j$ coupling schemes, respectively.
}
\end{center}
\end{figure}

One may complain that
the energy levels in the $j$-$j$ coupling scheme do not
agree with those of the $LS$ coupling scheme
at the first glance.
In order to reply to this complaint,
let us directly diagonalize $H_{\rm f}$ by changing $U$ and $\lambda$.
Here it is convenient to introduce a new parameter to connect the
$LS$ and $j$-$j$ coupling schemes as
\begin{equation}
 k=\frac{\lambda/|W|}{U/|W|+\lambda/|W|},
\end{equation}
where both $U$ and $\lambda$ are very large compared with
$|W|$, since we are considering the actual $f$-electron compound.
Note that $k$=0 and 1 are corresponding to the limits of
$\lambda/U$=0 and $\lambda/U$=$\infty$, respectively.
Then, we can control the change of two schemes by one parameter $k$.

In Figs.~2(a)-(d), we show the energy levels of $H_{\rm f}$
for several values of $k$ with both $\lambda$ and $U$
larger than $|W|$.
Racah parameters are set as $A/U$=10, $B/U$=0.3, $C/U$=0.1,
and $D/U$=0.05 in the unit of $U$.
As described above, the CEF potential is always small and 
here we set $W/U$=$-0.001$.
In Fig.~2(a), results for $k$=0.1 are shown.
In this case, $\lambda/U$=0.11 and
the condition $\lambda/|W|$$\gg$1 is still satisfied.
Without the spin-orbit interaction,
the ground-state level is expressed as $^{3}H$
with $S$=1 and $L$=5 due to Hund's rules.
When we increase $\lambda$,
the multiplet labeled by $J$ is well separated and
the ground-state level is specified by $J$=4,
as expected from the $LS$ coupling scheme.
Then, the energy levels in Fig.~2(a) are quite similar to
those of Fig.~1(a), since we are now in the region
where the $LS$ coupling scheme is appropriate.

Even when $\lambda$ is further increased and $k$ is equal to 0.5,
the structure of the energy levels is almost the same
as that of the $LS$ coupling scheme (not shown here).
However, when $k$ becomes 0.8, as shown in Fig.~2(b),
the energy level structure is found to be
deviated from that of the $LS$ coupling scheme.
Rather, it becomes similar to the energy level structure
of the $j$-$j$ coupling scheme.
To see the agreement with the $j$-$j$ coupling scheme, we consider
very large $\lambda$ to give $k$=0.99.
As shown in Fig.~2(d), we can observe the energy level structure
similar to Fig.~1(b).
Especially, the region of the $\Gamma_3$ ground state becomes
very narrow, as discussed later.
Thus, it is concluded that $H_{\rm f}$ correctly reproduces
the energy levels both for the $LS$ and $j$-$j$ coupling schemes.
We also stress that $H_{\rm f}$ provides correct results
for any value of $f$-electron number.

A crucial point is that the structure of energy levels is
continuously changed, as long as $\lambda$ and $U$ are
large compared with the CEF potential.
Namely, the states both in the $LS$ and $j$-$j$ coupling schemes
are continuously connected in the parameter space.
Thus, depending on the situation to consider the problem,
we are allowed to use the $LS$ or $j$-$j$ coupling scheme.
In order to clarify this point, we evaluate the overlap
$\langle \Psi |\Psi_{\rm LS}\rangle$,
where $|\Psi\rangle$ and $|\Psi_{\rm LS}\rangle$ are
the eigenstate of $H_{\rm f}$ and that in the $LS$ coupling scheme,
respectively.
In Fig.~2(d), we show the overlap for the case of $\Gamma_1$
ground state for $x$=$-0.1$, $y$=0.3, and $W/U$=$-0.001$.
For $k$=0, $\langle \Psi |\Psi_{\rm LS}\rangle$=1
due to the definition.
The overlap is gradually decreased with the increase of $k$,
but it smoothly converges to the value at $k$=1, i.e.,
the $j$-$j$ coupling scheme.
Note that the overlap between the eigenstates of the
$LS$ and $j$-$j$ coupling schemes is as large as 0.865,
which seems to be larger than readers may naively anticipated
from the clear difference between Figs.~1(a) and (b).
It is not surprising, if we are based on the principle of
adiabatic continuation, since the eigenstates of the
$LS$ and $j$-$j$ coupling schemes are continuously connected.

Remark that we can observe the common structure around
at the value of $x$, in which singlet and triplet
ground states are interchanged.
Namely, essential point of the singlet-triplet crossing
can be captured both in the two schemes, which will be important
in the following discussion.
However, the $\Gamma_3$ non-Kramers doublet cannot be the ground state
in the $j$-$j$ coupling scheme, since the doublet in the $J$=4 nontet
is composed of degenerate two singlets
formed by $\Gamma_7$ and $\Gamma_8$ electrons.
As easily understood, such singlets are energetically penalized
by the Hund's rule interaction and the energy for $\Gamma_4$ triplet
composed of $\Gamma_7$ and $\Gamma_8$ electrons is
always lower than that of the singlets.
Thus, in the $j$-$j$ coupling scheme, $\Gamma_3$ non-Kramers doublet
does not appear as the ground state except for $x$=0.
Of course, if $j$=7/2 octet is explicitly included and $\lambda$
is kept finite, it is possible to reproduce $\Gamma_3$ doublet.
Namely, taking account of the effect of $j$=7/2 octet is
equivalent to consider the local $f$-electron term $H_{\rm f}$,
as we have done in this subsection.
However, if we expand the Hilbelt space so as to include both
$j$=5/2 sextet and $j$=7/2 octet, we lose the advantage of
the $j$-$j$ coupling scheme considering only $j$=5/2 sextet.

One may claim that it is possible to reproduce the result of
the $LS$ coupling scheme even within the $j$-$j$ coupling scheme,
just by considering that the CEF potential for $J$=4 in
the $LS$ coupling scheme also works on the $J$=4 $f^2$-states
composed of a couple of $f$ electrons
among $j$=5/2 sextet.
However, such a procedure is $not$ allowed.
The reasons are as follows.
First it should be noted that the CEF potential
is $not$ determined only by the value of $J$.
For instance, as we will see later, the results of the energy
levels for $n$=7 and 13 are apparently different, even though both
of the ground-state multiplets are characterized by $J$=7/2,
since the CEF potential depends also on the values of $L$ and $S$.
Note that for $n$=7, $S$=7/2 and $L$=0, while for $n$=13,
$L$=3 and $S$=1/2.
For the case of $n$=2, even if the $f^2$-state is characterized
by $J$=4 in the $j$-$j$ coupling scheme,
we cannot simply validate the application of the
CEF potential in the $LS$ coupling scheme to the $J$=4
$f^2$-state in the $j$-$j$ coupling scheme.

Second we should note again that
the CEF effect appears only as a one-electron potential.
The CEF potential working on the two-electron state should be
given by the superimposition of the one-electron potential.
Thus, when we use the basis which diagonalizes the spin-orbit interaction,
it is necessary to consider that the CEF potential should work
on the state labeled by the $z$-component of $j$.
This is the only way to define the CEF potential in the $j$-$j$
coupling scheme, even though the $\Gamma_3$ non-Kramers doublet
is not reproduced.
As mentioned in the above paragraph, if $j$=7/2 octet
is included in addition to $j$=5/2 sextet, it is possible to
reproduce the results of the $LS$ coupling scheme including
the non-Kramers doublet.

\subsection{CEF schemes for filled skutterudites}

In the previous subsection, we have examined the local $f$-electronic
states in comparison with the results of the $LS$ and $j$-$j$ coupling
schemes for $n$=2.
In this subsection, we change the $f$-electron number from $n$=1 to 13,
in order to obtain the local $f$-electron states corresponding
from Ce- to Yb-based filled skutterudites.
Note that the results in this subsection will be comparable
with the experimental ones in the high-temperature region,
which is equivalent to the atomic limit.
In the calculation, we set the parameters as $A/U$=10, $B/U$=0.3,
$C/U$=0.1, $D/U$=0.05, $\lambda/U$=0.4, and $W/U$=$-0.001$
in the energy unit of $U$.
As already mentioned, we may consider $U$$\sim$1eV,
indicating that $\lambda$$\sim$0.4eV and $|W|$$\sim$1meV,
which are realistic values for rare-earth ions.
One of CEF parameters $y$ is also fixed as $y$=0.3 and we evaluate
the eigen energies of $H_{\rm f}$ as functions of $x$.
Note also that for each $n$, the energy is appropriately shifted
for the easy comparison.

\begin{figure}[t]
\begin{center}
\includegraphics[width=7truecm]{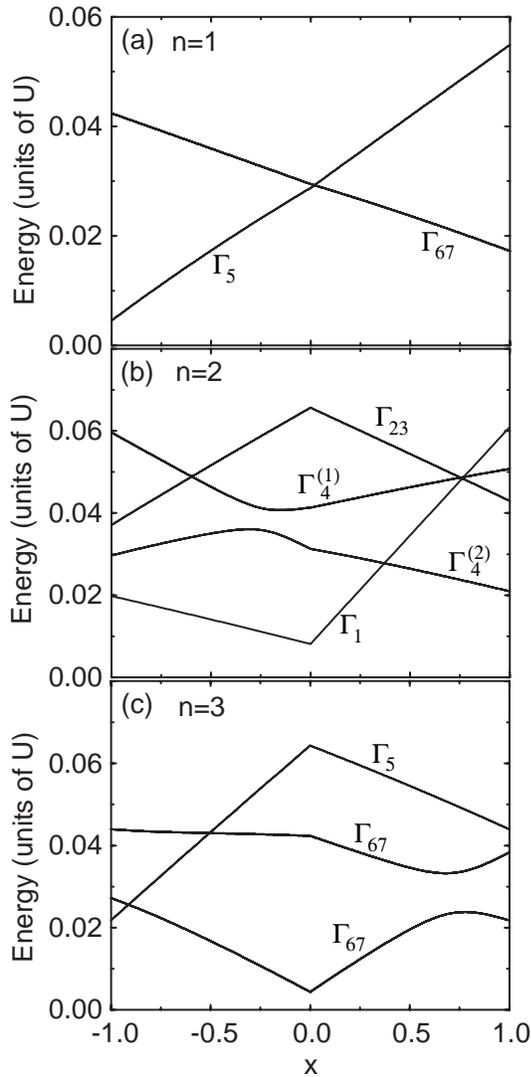}
\caption{Eigen energies of $H_{\rm f}$ as functions of $x$ for
(a) $n$=1, (b) $n$=2, and (c) $n$=3.
Other parameters are set as $A/U$=10, $B/U$=0.3, $C/U$=0.1, $D/U$=0.05,
$\lambda/U$=0.4, $W/U$=$-0.001$, and $y$=0.3.}
\end{center}
\end{figure}

\begin{figure}[t]
\begin{center}
\includegraphics[width=7truecm]{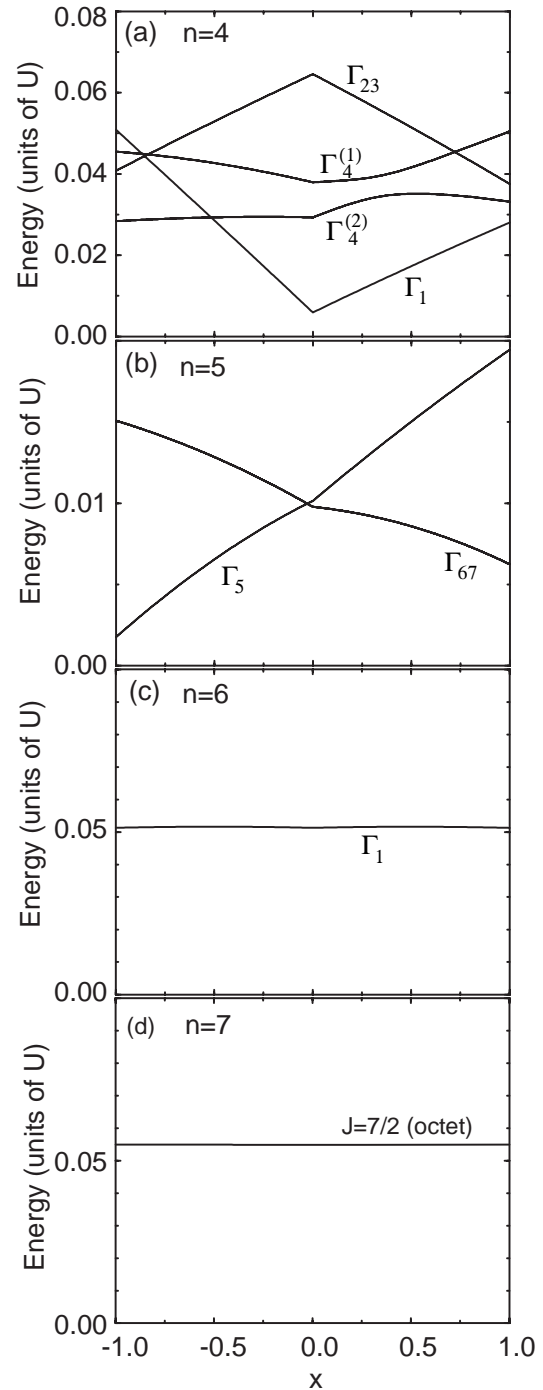}
\caption{Eigen energies of $H_{\rm f}$ as functions of $x$ for
(a) $n$=4, (b) $n$=5, (c) $n$=6, and (d) $n$=7.
Other parameters are the same as those in Fig.~3.}
\end{center}
\end{figure}

In Fig.~3, we show the results for $n$=1, 2, and 3, corresponding
to Ce$^{3+}$, Pr$^{3+}$, and Nd$^{3+}$ ions, respectively.
For $n$=1, as shown in Fig.~3(a), we observe two curves,
$\Gamma_5$ doublet and $\Gamma_{67}$ quartet.
Note that $\Gamma_5$ and $\Gamma_{67}$ in $T_{\rm h}$ correspond to
$\Gamma_7$ and $\Gamma_8$ in $O_{\rm h}$, respectively.\cite{Takegahara}
In order to determine $x$ for filled skutterudites, let us move
to the case of $n$=2. In Fig.~3(b), for $n$=2,
we can observe four eigen states, as described above.
Since it has been confirmed experimentally that the ground state is
$\Gamma_1$ and the first excited state is $\Gamma_4^{(2)}$ with small
excitation energy, it is quite reasonable to set $x$ around at 0.37.
Within a simple point-charge model, it is possible to fix
the CEF potential irrespective of $n$.
Thus, when we move back to the case of $n$=1, the ground state is
$\Gamma_{67}$ quartet.
When we turn our attention to Fig.~3(c) for the result of $n$=3,
in the region of $x$$>$0, the ground state is $\Gamma_{67}$ quartet,
consistent with the experimental results for
elastic constant\cite{Nakanishi} and
magnetic entropy\cite{Torikachvili} of NdFe$_4$P$_{12}$.

\begin{figure}[t]
\begin{center}
\includegraphics[width=7truecm]{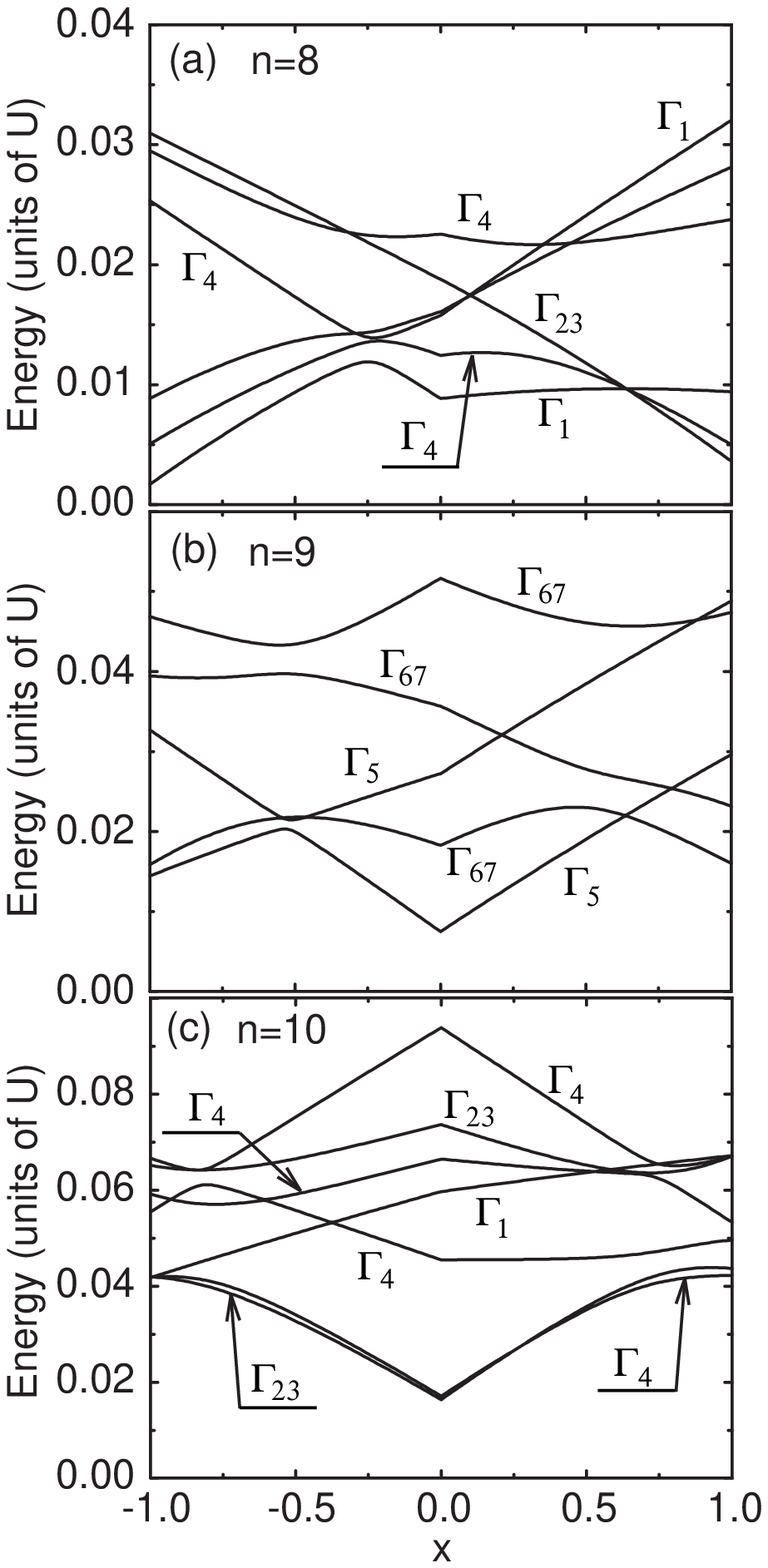}
\caption{Eigen energies of $H_{\rm f}$ as functions of $x$ for
(a) $n$=8, (b) $n$=9, and (c) $n$=10.
Other parameters are the same as those in Fig.~3.}
\end{center}
\end{figure}

Now let us see the results for $n$=4, 5, and 6.
The case of $n$=4 corresponds to Pm$^{3+}$, but Pm is an artificially
prepared element. Thus, we show the result only from the theoretical
interest. Since the multiplet including the ground state is
characterized by $J$=4 for the case of $n$=4, as shown in Fig.~4(a),
the structure of the eigen energies is essentially the same as that
for $n$=2. Note that the results of $x$$<$0 for $n$=2 appears
in the region of $x$$>$0 for $n$=4. Then, the ground state for $n$=4
should be $\Gamma_1$ singlet for $x$$>$0.
The case of $n$=5 is corresponding to Sm$^{3+}$.
As shown in Fig.~4(b), the ground state is $\Gamma_{67}$ quartet
for $x$$>$0, consistent with the experimental results of
elastic constant\cite{Yoshizawa} and magnetic entropy\cite{Matsuhira}
for SmRu$_4$P$_{12}$.
For $n$=6, the total angular momentum $J$ becomes zero,
since $S$=$L$=3 in this case. Thus, the singlet ground state appears
without the effect of CEF potential. The excitation energy is in the
order of 0.1, which will be about 1000K in the present energy unit.
The case of $n$=6 corresponds to trivalent Eu$^{3+}$,
but in the filled skutterudite structure,
significant contributions from Eu$^{2+}$ have been found in 
the measurement of magnetic susceptibility\cite{Sekine2} and 
$^{151}$Eu M\"ossbauer experiment.\cite{Indoh}
Divalent Eu$^{2+}$ ion has $f^7$ configuration, which
is discussed in the following.

At half-filling ($n$=7) corresponding to Eu$^{2+}$ or Gd$^{3+}$,
the result becomes quite simple.
Due to the Hund's rule coupling, the total spin $S$ is equal to 7/2,
while the angular momentum $L$ is zero.
Then, the total angular momentum $J$ is given by $J$=$S$=7/2, but
no effect of CEF potential occurs, leading to the $J$=7/2 octet
which is flat irrespective of $x$, as shown in Fig.~4(d).
As is well known in the theory for CEF,
since the CEF potential works only on the electron charge
distribution, the case of $L$=0 is not affected from the outset.
Note that the excitation energy in the present calculation
is estimated as 5eV.

\begin{figure}[t]
\begin{center}
\includegraphics[width=7truecm]{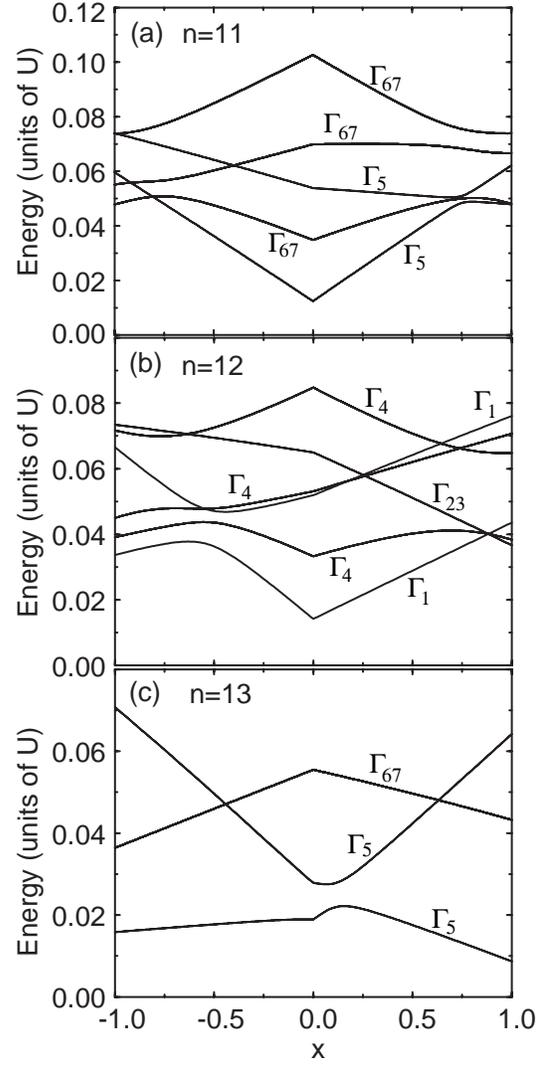}
\caption{Eigen energies of $H_{\rm f}$ as functions of $x$ for
(a) $n$=11, (b) $n$=12, and (c) $n$=13.
Other parameters are the same as those in Fig.~3.}
\end{center}
\end{figure}

Now we will move to the situation for $n$$>$7,
corresponding to heavy lanthanide ions.
In Fig.~5, we show the results for $n$=8, 9, and 10.
The case of $n$=8 corresponds to Tb$^{3+}$.
As shown in Fig.~5(a), for $n$=8,
the ground state is $\Gamma_1$ singlet around at $x$$\sim$0.37.
However, experimental results on magnetic susceptibility for
TbRu$_4$P$_{12}$ have suggested two successive magnetic transitions
at 20K and 10K.\cite{Sekine3}
As long as we fix $x$=0.37, the present CEF calculation indicates
the $\Gamma_1$ singlet ground state.
To explain the discrepancy, for instance, we may consider the
deviation from the ideal point-charge picture.
In particular, due to the difference in the ion radius,
the CEF potential can be different from material to material
and the values of CEF parameters may be changed.
If we slightly increase the value of $x$,
the ground state turns to be $\Gamma_4$ triplet,
leading to the magnetic properties at low temperatures
observed in Tb-based filled skutterudites.

In Fig.~5(b), the result for $n$=9 corresponding to Dy$^{3+}$ ion
is shown. Around at $x$=0.37, the ground state is $\Gamma_5$ doublet,
which is $\Gamma_7$ doublet in the term of $O_{\rm h}$ symmetry.
Thus, the ground state should be magnetic, but the effect of
quadrupole may appear in Dy-based filled skutterudites,
if the CEF potential is deviated from the present one.
In Fig.~5(c), we show the result of $n$=10, corresponding to Ho$^{3+}$
ion. In this situation, around at $x$$\sim$0.37, the ground state
is $\Gamma_4$ triplet, while $\Gamma_{23}$ non-Kramers doublet is
the first excited state with very small energy difference,
less than 10K in the present energy unit.
In addition, as mentioned above, $x$ may be changed from 0.37,
which is the value determined for Pr-based filled skutterudite.
Thus, the ground state of Ho-based filled skutterudite is believed
to have magnetic nature, but the effect of quadrupole degree of freedom
of the $\Gamma_{23}$ doublet may appear as an anomaly
in physical quantities such as elastic constant.

Finally in Fig.~6, we show the results for $n$=11, 12, and 13,
corresponding to Er$^{3+}$, Tm$^{3+}$, and Yb$^{3+}$ ions.
As observed in Figs.~6(a)-(c), the ground state is $\Gamma_5$
Kramers doublet for $n$=11 or 13, while $\Gamma_1$ singlet for $n$=12
around at $x$$\sim$0.37.
Note that these states are not drastically changed in the region
of 0$<$$x$$<$0.8, indicating the stable ground states
for the filled skutterudites with $n$=11, 12, and 13.

%
%
\section{Multiorbital Anderson Model}

In the previous section, we have discussed in detail the local
$f$-electron states for $n$=1$\sim$13.
However, in order to consider the low-energy state,
it is necessary to include the itinerant effect of $f$ electron.
Although the actual material should be described by the periodic system,
it is a hard task to treat such a periodic model by keeping seven
$f$-orbitals. Thus, in this section, we consider the impurity Anderson
model, by taking into account of the hybridization between $f$ and
conduction electron states. The symmetry of the conduction band is
important to understand correctly the low-energy properties of
filled skutterudite compounds.

\subsection{Hamiltonian}

The Anderson Hamiltonian is written by
\begin{eqnarray}
  \label{AndersonModel}
  H = \sum_{\mib{k}\sigma}
  \varepsilon_{\mib{k}} c_{\mib{k}\sigma}^{\dag} c_{\mib{k}\sigma}
  + \sum_{\mib{k}\sigma m}
  (V_{m} c_{\mib{k}\sigma}^{\dag}f_{m\sigma}+{\rm h.c.})
  + H_{\rm f},
\end{eqnarray}
where $\varepsilon_{\mib{k}}$ is the dispersion of conduction electron,
$c_{\mib{k}\sigma}$ is the annihilation operator for conduction
electron with momentum $\mib{k}$ and spin $\sigma$, and
$V_{m}$ is the hybridization between conduction and $f$ electrons.
The local $f$-electron term $H_{\rm f}$ is already given in the
previous section.
In the filled skutterudites, the conduction band is given by $a_{\rm u}$,
constructed from $p$-orbitals of pnictogen.\cite{Harima}
Note that the hybridization occurs between the states with the same
symmetry. Since the $a_{\rm u}$ conduction band has xyz symmetry, we set
\begin{eqnarray}
 V_m=\left\{
 \begin{array}{ll}
    V & {\rm for}~m=2, \\
   -V & {\rm for}~m=-2, \\
    0 &{\rm otherwise.}
 \end{array}
 \right.
\end{eqnarray}
Throughout this section, the energy unit is taken as $D_0$,
which is half of the bandwidth of the conduction band.
From the band-structure calculation, the bandwidth is
estimated as 2.7 eV in PrRu$_4$P$_{12}$,\cite{Harima1}
indicating that $D_0$=1.35 eV.
Note that in order to adjust the local $f$-electron number,
we change the chemical potential for each $n$.

\subsection{Method}

In this paper, the Anderson model is analyzed by using the numerical
renormalization group (NRG) technique.\cite{Wilson,NRG}
In the NRG calculations, in order to consider efficiently the
conduction electrons near the Fermi energy, the momentum space
is logarithmically discretized and the conduction electron states
are characterized by ``shell'' labeled by $N$.
The shell of $N$=0 denotes an impurity site including $f$ electrons.
The Hamiltonian is transformed into the recursion form as
\begin{eqnarray}
  H_{N+1} = \sqrt{\Lambda}H_N+t_N \sum_\sigma
  (c_{N\sigma}^{\dag}c_{N+1\sigma}+c_{N+1\sigma}^{\dag}c_{N\sigma}),
\end{eqnarray}
where $\Lambda$ is a parameter for logarithmic discretization,
$c_{N\sigma}$ denotes the annihilation operator of conduction electron
in the $N$-shell, and $t_N$ is ``hopping'' of electron between $N$-
and $(N+1)$-shells, given by
\begin{eqnarray}
  t_N=\frac{(1+\Lambda^{-1})(1-\Lambda^{-N-1})}
  {2\sqrt{(1-\Lambda^{-2N-1})(1-\Lambda^{-2N-3})}}.
\end{eqnarray}
The initial term $H_0$ is given by
\begin{eqnarray}
  H_0=\Lambda^{-1/2}[H_{\rm f} + \sum_{m\sigma}
  V_{m}(c_{0\sigma}^{\dag}f_{m\sigma}+f_{m\sigma}^{\dag}c_{0\sigma})].
\end{eqnarray}
In this paper, $\Lambda$ is set as $5$ and we keep $4000$
low-energy states for each renormalization step.

In order to see magnetic properties, we evaluate
the magnetic susceptibility of $f$-electron, defined by
\begin{eqnarray}
   \chi \!=\! \frac{1}{T} \lim_{N \rightarrow \infty}
   \Biggl[ \frac{{\rm Tr} M_{z,N}^2 e^{-H_N/T}}{{\rm Tr} e^{-H_N/T}}
   - \frac{{\rm Tr} S_{z,N}^2 e^{-H^0_N/T}}{{\rm Tr} e^{-H^0_N/T}}
   \Biggr],
\end{eqnarray}
where $T$ is a logarithmic temperature given by $T$=$\Lambda^{-(N-1)/2}$
in the NRG calculation, $H_N^0$ is the Hamiltonian without $H_{\rm f}$,
and magnetic moment is defined by
\begin{eqnarray}
   M_{z,N}=-\mu_{\rm B}\sum_{m,\sigma}(m+g_s\sigma/2)
   f_{m\sigma}^{\dag}f_{m\sigma}+S_{z,N}.
\end{eqnarray}
Here $\mu_{\rm B}$ is the Bohr magneton, $g_s$=2, and
\begin{eqnarray}
  S_{z,N}=-g_s\mu_{\rm B}\sum_{n=0}^{N}\sum_{\sigma}(\sigma/2)
   c_{n\sigma}^{\dag}c_{n\sigma}.
\end{eqnarray}
The free energy $F$ for $f$ electron is evaluated by
\begin{eqnarray}
   F \!=\! -T \lim_{N \rightarrow \infty}
   \Biggl[ \ln {\rm Tr} e^{-H_N/T}
   \!-\! \ln {\rm Tr} e^{-H_N^0/T} \Biggr].
\end{eqnarray}
The entropy $S$ is obtained by $S$=$-\partial F/\partial T$.

\begin{figure}[t]
\begin{center}
\includegraphics[width=7truecm]{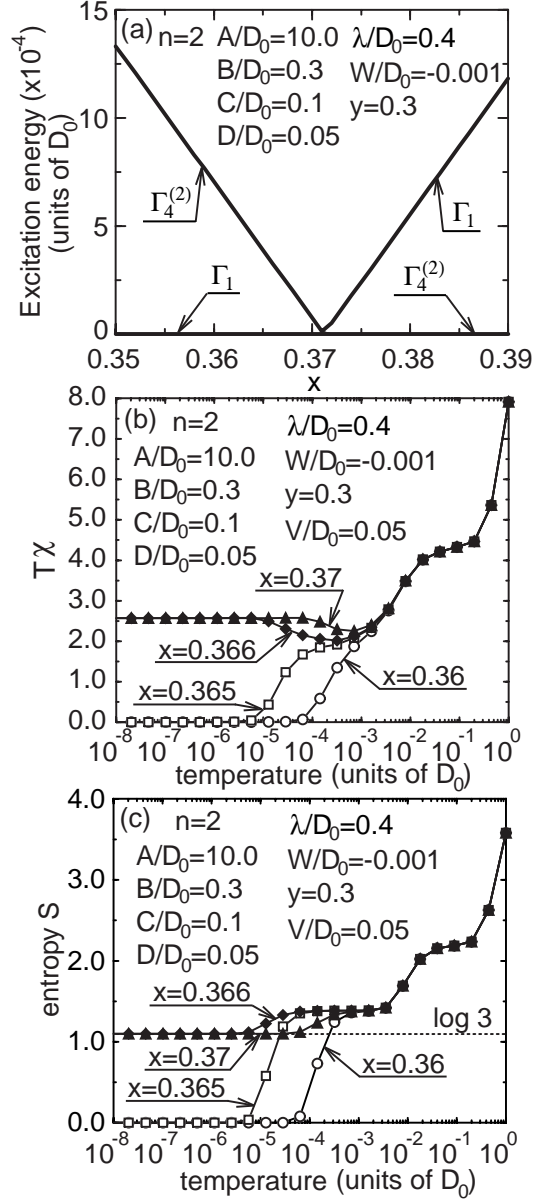}
\caption{(a) Excitation energy vs. $x$ around at $x$$\sim$0.37
for $n$=2. (b) Magnetic susceptibility and (c) entropy
of $f$ electron vs. $T$ for $n$=2 and $x$=0.36, 0.365, 0.366,
and 0.37.}
\end{center}
\end{figure}

\subsection{Result for $n$=2}

First let us briefly review the results for $n$=2,\cite{Hotta4,Hotta5}
corresponding to the Pr-based filled skutterudites.
As mentioned in the previous section, it has been confirmed experimentally
that the ground state is $\Gamma_1$ and the first excited state is
$\Gamma_4^{(2)}$ with small excitation energy
for Pr-based filled skutterudites.
In Fig.~7(a), we show the excitation energy for $n$=2 around at $x$=0.37
in a magnified scale. Then, we consider the region of
0.36$<$$x$$<$0.37 in the following.

In Fig.~7(b), we show calculated results of $T\chi$ for $n$=2 and
several values of $x$ between 0.36 and 0.37. For $x$$<$0.365, $T\chi$
becomes zero for small $T$, while for $x$$>$0.366, $T\chi$ takes constant
at low temperatures.\cite{Hotta4}
As shown in Fig.~7(c) for the calculated results of entropy $S$,
we obtain $\log 3$ as residual entropy for $x$=0.366 and 0.37,
indicating that local triplet remains in the low-temperature region.
Of course, the local moment is eventually suppressed at low temperatures
in actuality. For instance, there should exist the hybridization
with conduction bands other than $a_{\rm u}$.
What we emphasize here is that magnetic fluctuations still remain
at relatively low temperatures even when $\Gamma_1$ is the ground state,
if $\Gamma_4^{(2)}$ is the excited state with small excitation energy.
In the following calculation, we set $x$=0.366.

We remark that the transition between magnetic
and non-magnetic states is governed by the exchange
interaction $J_{\rm cf}$ between $f$ electrons and the conduction band,
expressed as $J_{\rm cf}=V^2/\delta \! E$,
where $\delta \! E$ denotes the energy difference between
the $f^2$ and $f^3$ (or $f^1$) lowest-energy states.
In fact, it has been found that the boundary curve
between magnetic and non-magnetic phases
is proportional to $V^2$,\cite{Hotta5}
suggesting that the dominant energy scale should be $J_{\rm cf}$
for the appearance of magnetic fluctuations.
Note that the transition between magnetic and non-magnetic
states seems to be abrupt, not gradual as observed in the usual
Kondo system.
In the Kondo problem, the local moment is suppressed
$only$ by hybridization with conduction electrons,
but in the present case, a singlet ground state is also
obtained through local level crossing due to the CEF potential,
in addition to the hybridization process.
Furthermore, localized orbitals exist in the present model,
as will be discussed in the next subsection.
Namely, the magnetic moments of the $f$ orbitals hybridized with
the conduction band are suppressed as in the Kondo effect, while
the moments of the localized orbitals are not screened.
An abrupt change in the magnetic properties is caused by
the duality of the $f$ orbitals in combination with 
the level crossing effect due to the CEF potential.

\subsection{Reconsideration in a $j$-$j$ coupling scheme}

In order to understand the physical meaning of the result for $n$=2,
it is useful to express the $f$-electron state with
xyz symmetry in the $j$-$j$ coupling scheme as
\begin{equation}
 |{\rm xyz},\sigma \rangle=
 \sqrt{4/7}|7/2, \Gamma_7,{\tilde \sigma} \rangle-
 \sqrt{3/7} |5/2, \Gamma_7,{\tilde \sigma} \rangle,
\end{equation}
where $|j,\Gamma,{\tilde \sigma} \rangle$
denotes the state in the $j$-$j$ coupling scheme
with total angular momentum $j$, irreducible representation
$\Gamma$ for the $O_{\rm h}$ point group,
and pseudospin ${\tilde \sigma}$.
We note that in the $j$-$j$ coupling scheme,
we use the irreducible representation of $O_{\rm h}$,
since the CEF potential for one $f$-electron state
in the $j$=5/2 sextet is
the same for $O_{\rm h}$ and $T_{\rm h}$.
Note also that there is one-to-one correspondence
between $\sigma$ and ${\tilde \sigma}$.
The above relation indicates that
only the $\Gamma_7$ state is hybridized with
$a_{\rm u}$ conduction band states with xyz symmetry,
leading to the Kondo effect,
while the $\Gamma_8$ electrons are localized
and become the source of local fluctuations at low temperatures.
It is an important issue of filled skutterudite structures
that the nature of the $f$ electrons can clearly be distinguished
as itinerant $\Gamma_7$ or localized $\Gamma_8$.
This point provides a possible explanation for the heavy-fermion
phenomenon occurring in $f^2$-electron systems such as
Pr-based filled skutterudites.

In the $j$-$j$ coupling scheme,
we generally assume the large spin-orbit coupling, indicating that
$j$=7/2 octet is discarded and only $j$=5/2 sextet is taken into account.
We also note that the conduction band has $\Gamma_7$ symmetry
in the filled skutterudite compounds.
Then, the Anderson Hamiltonian in the $j$-$j$ coupling scheme is written as
\begin{eqnarray}
  H = \sum_{\mib{k}\sigma}
  E_{\mib{k}} a_{\mib{k}\sigma}^{\dag} a_{\mib{k}\sigma}
  +\sum_{\mib{k}\sigma}
  (V a_{\mib{k}\sigma}^{\dag}f_{{\bf i}{\rm c}\sigma}+{\rm h.c.})
  +H_{\bf i},
\end{eqnarray}
where $E_{\mib{k}}$ is the dispersion of $\Gamma_7$ conduction electron,
the energy unit in this subsection is also taken as $D_0$ which is
half of the bandwidth of the conduction band,
$a_{\mib{k}\sigma}$ is the annihilation
operator for $\Gamma_7$ conduction electron with momentum $\mib{k}$ and
pseudospin $\sigma$, $f_{{\bf i}\gamma\sigma}$ is the annihilation
operator for $j$=5/2 $f$-electron on the impurity site ${\bf i}$
with spin $\sigma$ and ``orbital'' $\gamma$, $V$ is the hybridization
between conduction and $f$-electrons with $\Gamma_7$ symmetry, and
$H_{\bf i}$ denotes the local $f$-electron term at site ${\bf i}$.
Note that the orbital index $\gamma$ is introduced to distinguish
three kinds of the Kramers doublets, two $\Gamma_8$ and one $\Gamma_7$.
Here ``a'' and ``b'' denote two $\Gamma_8$ and ``c'' indicates $\Gamma_7$.

The local $f$-electron term $H_{\bf i}$ in the $j$-$j$ coupling scheme
is given as\cite{Hotta1}
\begin{eqnarray}
  H_{\bf i} &=& \sum_{\gamma,\sigma} {\tilde B}_{\gamma}
  f_{{\bf i}\gamma\sigma}^{\dag}f_{{\bf i}\gamma\sigma}+
  (1/2) \sum_{\gamma_1 \sim \gamma_4}\sum_{\sigma_1, \sigma_2}
  {\tilde I}^{\sigma_1,\sigma_2}_{\gamma_1, \gamma_2, \gamma_3, \gamma_4}
  \nonumber \\
  &\times& f_{{\bf i}\gamma_1\sigma_1}^{\dag}
  f_{{\bf i}\gamma_2\sigma_2}^{\dag}
  f_{{\bf i}\gamma_3\sigma_2}f_{{\bf i}\gamma_4\sigma_1},
\end{eqnarray}
where ${\tilde B}_{\gamma}$ is the CEF potential for $j$=5/2.
Since the CEF potential is already diagonalized, it is convenient
to introduce a level splitting between $\Gamma_7$ and $\Gamma_8$ as
\begin{equation}
  \Delta={\tilde B}_{\Gamma_8}-{\tilde B}_{\Gamma_7}=6Wx.
\end{equation}
The Coulomb integral
${\tilde I}^{\sigma_1,\sigma_2}_{\gamma_1,\gamma_2,\gamma_3,\gamma_4}$
in the $j$-$j$ coupling scheme is expressed by using other
Racah parameters, $E_0$, $E_1$, and $E_2$.\cite{Hotta1}

\begin{figure}[t]
\begin{center}
\includegraphics[width=7truecm]{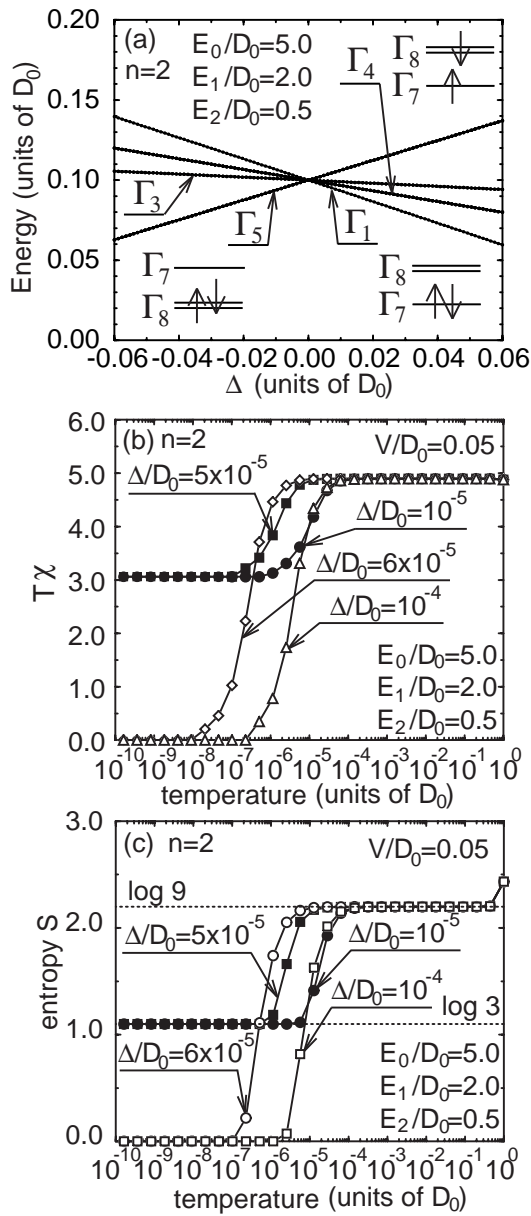}
\caption{(a) Eigen energies of $H_{\bf i}$ for $n$=2.
Insets denote $f$-electron configurations
for $\Gamma_1$, $\Gamma_4$, and $\Gamma_5$ states,
in which two $f$ electrons
are accommodated in the $f^1$-energy levels.
(b) Susceptibility and (c) entropy of $f$ electron vs. $T$
for $n$=2 based on the Anderson model in the $j$-$j$ coupling scheme
for various values of $\Delta$.}
\end{center}
\end{figure}

First let us discuss the local $f$-electronic properties of $H_{\bf i}$,
when we change the level splitting $\Delta$, although the essential
point has been already discussed in Sec.~2.
As shown in Fig.~8(a), the ground state is $\Gamma_5$ triplet
for $\Delta$$<$0, while it is $\Gamma_1$ singlet for $\Delta$$>$0.
Note that for positive $\Delta$, the triplet excited state
is $\Gamma_4$, composed of $\Gamma_7$ and $\Gamma_8$ electrons.
Note again that we use the irreducible representation of $O_{\rm h}$,
since there is no difference between $O_{\rm h}$ and $T_{\rm h}$
in the $j$-$j$ coupling scheme.
We again apply the NRG method
to the Anderson model in the $j$-$j$ coupling scheme.
In Figs.~8(b) and (c), we show the results of $T\chi$ and
the entropy $S$ of $f$ electrons for several values of $\Delta$.
Racah parameters are set as $E_0/D_0$=5, $E_1/D_0$=2, and $E_2/D_0$=0.5,
which are appropriate for rare-earth compounds.
The hybridization is set as $V/D_0$=0.05.
For high temperatures, since all the $J$=4 states contribute to
the $f$-electronic properties, we obtain the entropy of $\log 9$.
For $\Delta/D_0$$>$$6\times10^{-5}$, both susceptibility and entropy
eventually go to zero in the low-temperature region, while
for $\Delta/D_0$$<$$5\times10^{-5}$, magnetic fluctuations remain
significant at low temperatures. Note that the residual entropy is
$\log 3$ for $\Delta/D_0$$<$$5\times10^{-5}$.

Now we consider the reason why the essential magnetic features
of Pr-based filled skutterudites can be captured as described
in the $j$-$j$ coupling scheme.
As shown in Fig.~8(a), when the ground state is a $\Gamma_1$
singlet, there are two triplet excited states,
$\Gamma_4$ and $\Gamma_5$.
The $\Gamma_5$ triplet is composed of a pair of $\Gamma_8$
electrons, while the $\Gamma_4$ triplet is composed of one
$\Gamma_7$ and one $\Gamma_8$ electron.
Since $\Gamma_8$ electrons do $not$ hybridize with $\Gamma_7$
conduction electrons, the $\Gamma_5$ triplet can survive.
Note that $\Gamma_4$ triplets in $T_{\rm h}$ are given by
the mixtures of $\Gamma_4$ and $\Gamma_5$ in $O_{\rm h}$.
Such a mixing is not included in the $j$-$j$ coupling scheme,
but there exists a $\Gamma_5$ excited state.
Thus, the local triplet still remains, as long as
the excitation energy is smaller than $J_{\rm cf}$.

Summarizing this subsection,
the low-temperature $f$-electron properties can be
well captured even in the Anderson model constructed from
the $j$-$j$ coupling scheme, in the parameter region in which singlet
and triplet ground states are interchanged.
In particular, even though the ground state is $\Gamma_1$ singlet,
the magnetic fluctuations survive in the low-temperature region,
when the triplet excited state exists with very small excitation energy.
Thus, the $j$-$j$ coupling scheme is applicable to investigate the
microscopic $f$-electron properties of rare-earth filled skutterudites.

\begin{figure}[t]
\begin{center}
\includegraphics[width=7truecm]{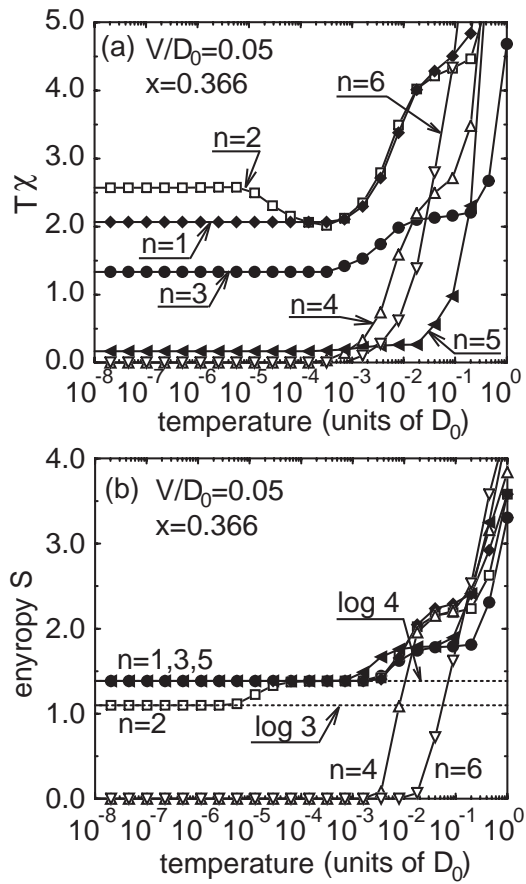}
\caption{(a) Magnetic susceptibility and (b) entropy
of $f$ electron vs. $T$ for $n$=1$\sim$6.
Racah parameters are set as $A/D_0$=10, $B/D_0$=0.3, $C/D_0$=0.1, and
$D/D_0$=0.05. Spin-orbit coupling and $c$-$f$ hybridization are
set as $\lambda/D_0$=0.4 and $V/D_0$=0.05, respectively.
As for CEF parameters, we set $W/D_0$=$-0.001$, $x$=0.366, and $y$=0.3.}
\end{center}
\end{figure}

\subsection{Results for $n$=1$\sim$13}

Thus far, we have focused on the case of $n$=2, but we can
further study the cases of $n$=1$\sim$13
based on the original Anderson model Eq.~(\ref{AndersonModel}).
In the point-charge picture, the CEF parameters do $not$
depend on $n$, since the CEF effect is due to electrostatic potential
from ligand ions surrounding the rare earth.
In the following, we fix $x$=0.366,
in which the local $f^2$ ground state is a $\Gamma_1$ singlet,
but for which we have found significant magnetic fluctuations at low
temperatures.

In Fig.~9(a), magnetic susceptibilities for $n$$<$7 are shown for
$x$=0.366. For $n$=2, 4, and 6, the ground state is $\Gamma_1$ singlet,
indicating that both magnetic and orbital fluctuations should be
rapidly suppressed with decreasing temperature.
However, for $n$=2, since the $\Gamma_4^{(2)}$ triplet is existing
with very small excitation energy, magnetic fluctuations significantly
remain, as mentioned in the previous subsection.
Note that for $n$=4 and 6, the excitation energy is large and $T\chi$
becomes zero immediately in the low-temperature region.
On the other hand, for $n$=1, 3, and 5, the ground state is
$\Gamma_{67}$ quartet, as confirmed from the residual entropy of
$\log 4$, as shown in Fig.~9(b).
Here we remark that the case of $n$=5 corresponds to Sm$^{3+}$ ion.
In NMR experiments for Sm-based filled skutterudites,\cite{Sm}
it has been suggested that magnetic fluctuations seem to appear in the
low-temperature region, consistent with the present numerical result.

\begin{figure}[t]
\begin{center}
\includegraphics[width=7truecm]{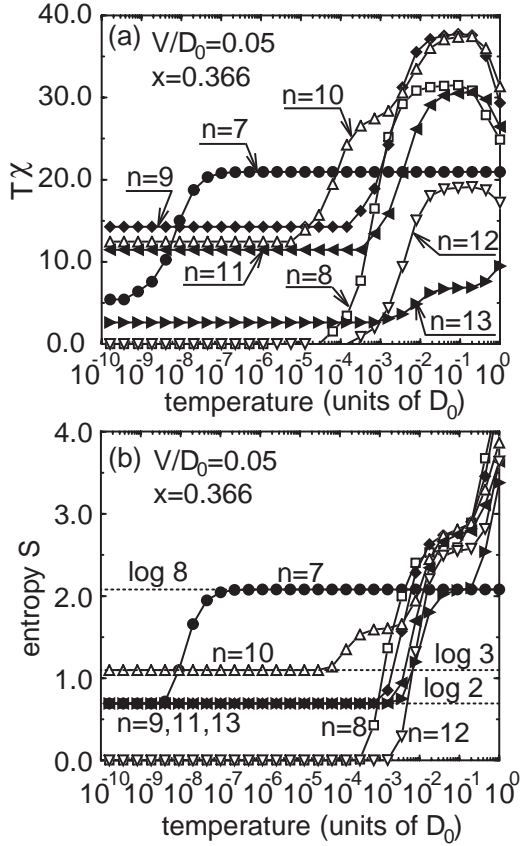}
\caption{(a) Magnetic susceptibility and (b) entropy
of $f$ electron vs. $T$ for $n$=7$\sim$12.
Racah parameters are set as $A/D_0$=10, $B/D_0$=0.3, $C/D_0$=0.1, and
$D/D_0$=0.05. Spin-orbit coupling and $c$-$f$ hybridization are
set as $\lambda/D_0$=0.4 and $V/D_0$=0.05, respectively.
As for CEF parameters, we set $W/D_0$=$-0.001$, $x$=0.366, and $y$=0.3.}
\end{center}
\end{figure}

Next we turn our attention to the cases of half-filling ($n$=7)
and more than half-filling ($n$$>$7).
In Fig.~10(a), we show the NRG results for susceptibility.
First of all, the absolute values of $\chi$ are much larger than
those for $n$$<$7, since the magnetic moment becomes large
for $n$$>$7 due to the Hund's rule interactions.
Typically, at half-filling, total spin $S$(=$J$) is equal to 7/2
and the Curie constant for an isolated ion is as large as
21 $\mu_{\rm B}^2/k_{\rm B}$.
In the wide temperature region, this value has been observed
in Fig.~10(a) for $n$=7, indicating that $S$=7/2 spin survives
at relatively low temperatures.
In fact, we clearly see the entropy of $\log 8$ from the octet of
$S$=7/2, as shown in Fig.~10(b).
In the extremely low-temperature region, however, eventually $S$=7/2
spin is screened and only $S$=1/2 remains, leading to the residual
entropy $\log 2$.

For the cases of $n$=8 and 12, as discussed in the previous subsection,
the ground state is $\Gamma_1$ singlet and thus, the susceptibility
rapidly goes to zero. Note that in the case of $n$=8,
when we change the value of $x$, the effect of the magnetic excited
state becomes significant, as in the case of $n$=2.
In order to focus on the case of Tb-based filled skutterudites,
it is necessary to analyze in detail the experimental results,
but it will be discussed elsewhere in future.
For $n$=9, 11, and 13, the local ground state is $\Gamma_5$ Kramers
doublet, which is the mixture of $\Gamma_6$ and $\Gamma_7$ of
$O_{\rm h}$.
Since $\Gamma_6$ state does not hybridize with the $a_{\rm u}$
conduction band, the magnetic moment from $\Gamma_6$ still persists
even in the low-temperature region.
In fact, we observe the residual entropy of $\log 2$ in these cases,
as shown in Fig.~10(b).

For $n$=10, the local ground state is $\Gamma_4$ triplet, but
it just remains as it is in the low-temperature region,
as observed in Figs.~10(a) and (b). This is understood as follows:
It is convenient to reconsider the states in the $j$-$j$ coupling scheme
and $O_{\rm h}$ symmetry. The $\Gamma_5$ triplet in the $O_{\rm h}$ symmetry
is composed of a couple of $\Gamma_8$ electrons, while
the $\Gamma_4$ triplet in the $O_{\rm h}$ symmetry is composed of
$\Gamma_7$ and $\Gamma_8$ electrons.
Namely, the $\Gamma_5$ triplet can survive, since $\Gamma_8$ electron
does not hybridize with $\Gamma_7$ conduction electron.
In the $T_{\rm h}$ symmetry, $\Gamma_4$ triplet states are always given
by the mixture of $\Gamma_4$ and $\Gamma_5$ triplets of $O_{\rm h}$ and
thus, the $\Gamma_5$ component still remains,
even after the hybridization with conduction electrons.

%
%
\section{Superconductivity of Pr-Based Filled Skutterudites}

In order to discuss the appearance of superconductivity
from a microscopic viewpoint, here we simply consider the Hubbard-like
model for $f$-electron systems,\cite{Hotta1}
in which $\Gamma_7$ electrons move around the system through
$f$-$f$ hopping in the present case.
Note that it is necessary to treat the periodic Anderson model
in order to consider simultaneously the formation of heavy quasi-particle
and the occurrence of superconductivity.
However, for the present purpose to discuss the nature of
superconducting pairing, we believe that the construction of
the Hubbard-like model for $f$-electron systems is useful.
The Hamiltonian is given by
\begin{eqnarray}
  \label{HubbardModel}
  H = -t \sum_{\langle {\bf i},{\bf j} \rangle \sigma}
  (f_{{\bf i}{\rm c}\sigma}^{\dag} f_{{\bf j}{\rm c}\sigma}+{\rm h.c.})
  + \sum_{\bf i} H_{\bf i},
\end{eqnarray}
where $t$ denotes an effective hopping amplitude of $\Gamma_7$ electron
between nearest neighbor sites, $\langle {\bf i},{\bf j} \rangle$,
in the bcc lattice.
In this section, $t$ is taken as a new energy unit.
We believe that
this Hamiltonian can be a canonical model to investigate superconductivity
of filled skutterudites from the microscopic viewpoint.

Now we evaluate superconducting pair susceptibility $P$.
For the purpose, we define the pair operator $\hat{\Phi}$ as
\begin{eqnarray}
  \hat{\Phi} = \sum_{{\bf i}\gamma\sigma{\bf i'}\gamma'\sigma'}
  \varphi_{{\bf i}\gamma\sigma{\bf i'}\gamma'\sigma'}
  \hat{\Phi}_{{\bf i}\gamma\sigma{\bf i'}\gamma'\sigma'},
\end{eqnarray}
where
$\hat{\Phi}_{{\bf i}\gamma\sigma{\bf i'}\gamma'\sigma'}=
f_{{\bf i}\gamma\sigma}f_{{\bf i'}\gamma'\sigma'}$
and $\varphi_{{\bf i}\gamma\sigma{\bf i'}\gamma'\sigma'}$
is the coefficient determined by the eigen state of
the maximum eigen value $P_{\rm max}$ of the matrix $P$, given by
\begin{eqnarray}
  P_{{\bf i}\gamma\sigma{\bf i'}\gamma'\sigma',{\bf j}\mu\nu{\bf j'}\mu'\nu'}
  \!=\! \int_0^{1/T} \! d\tau \langle 
  \hat{\Phi}_{{\bf i}\gamma\sigma{\bf i'}\gamma'\sigma'}(\tau)
  \hat{\Phi}^{\dag}_{{\bf j}\mu\nu{\bf j'}\mu'\nu'} \rangle.
\end{eqnarray}
Here $\hat{A}(\tau)$=$e^{H\tau}\hat{A}e^{-H\tau}$ for an operator $\hat{A}$.
Information on the symmetry of Cooper pair is included in $\varphi$.

\begin{figure}[t]
\begin{center}
\includegraphics[width=7truecm]{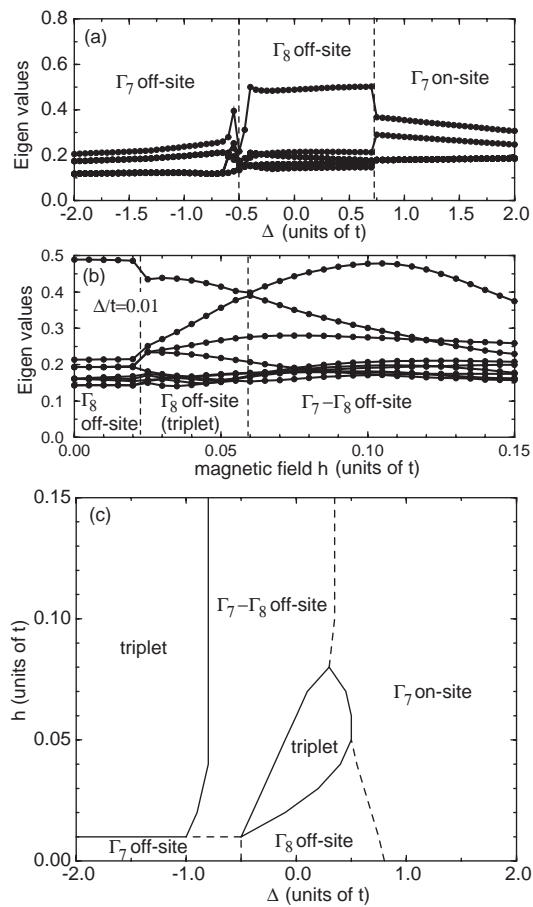}
\caption{(a) Eigen values of pair susceptibility
vs. $\Delta$ in the 2-site model for $T/t$=$10^{-3}$.
(b) Eigen values of $P$ as functions of magnetic field
$\Delta/t$=0.01 in the 2-site model.
(c) Phase diagram for superconducting pair in the 2-site model.
Solid curve indicate the boundary between singlet and triplet pairs,
while the broken curve denotes the boundary between two singlet pair
states.}
\end{center}
\end{figure}

In Fig.~11(a), eigen values of pair susceptibility are plotted as functions
of $\Delta$ for the 2-site case.
Since three orbitals are included at each site, it is difficult to enlarge
the system size for the evaluation of susceptibility, but we can grasp
some essential points of pair wavefunction in an unbiased manner.
First we observe that only in the region of $\Delta/t$$\approx$0,
$P_{\rm max}$ is significantly enhanced.
The corresponding eigen state indicates an off-site singlet pair of
$\Gamma_8$ electrons. For $\Delta/t$$\ll$1 and $\Delta/t$$\gg$1,
we obtain off-site and on-site singlet pair of $\Gamma_7$ electrons,
respectively. Here we emphasize a concept that the appearance of
superconductivity is controlled by $\Delta$.
When $|\Delta|$ is increased, pair susceptibility is suppressed
in any case, since insulating states are dominant.
For positive $\Delta$ with $\Delta/t$$\gg$1, $\Gamma_7$ is doubly
occupied, leading to charge-density-wave (CDW) state,
while for negative $\Delta$ with $|\Delta|/t$$\gg$1,
spin $S$=1 is formed by a couple of $\Gamma_8$ electrons,
leading to spin-density-wave (SDW) state.
When two different insulating phases compete with each other,
a metallic state appears in the competing region \cite{Takada}
and the superconductivity is expected to appear there.\cite{Hotta5}

Let us see what happens under the magnetic field.
We evaluate the pair susceptibility by adding the Zeeman term to
the Hubbard model, given by
\begin{equation}
  H_{\rm Z} = -g_{J} \mu_{\rm B} H \sum_{{\bf i},\mu}
  \mu \alpha_{{\bf i}\mu}^{\dag} \alpha_{{\bf i}\mu},
\end{equation}
where $g_{J}$ is the Land\'e's $g$-factor ($g_{J}$=6/7 for $j$=5/2),
$H$ is an applied magnetic field,
$\mu$ is the $z$-component of total angular momentum $j$=5/2,
and $\alpha_{{\bf i}\mu}$ is the annihilation operator
of $f$-electron labeled by $\mu$ at site ${\bf i}$.
For convenience, we introduce the non-dimensional
magnetic field as $h$=$g_{J} \mu_{\rm B} H/(k_{\rm B}t)$.
Note that in actual calculations, we use the $f$-electron basis
so as to diagonalized the CEF term.
As shown in Fig.~11(b), the state of $P_{\rm max}$ at $\Delta/t$=0.01
is found to have odd-parity triplet pair in the region
of a small magnetic field.
When we increase the magnetic field, there occurs
another singlet off-site pair state composed of
$\Gamma_7$ and $\Gamma_8$ electrons.

Based on the 2-site calculation results,
we depict the phase diagram
for superconducting pair state in the plane of $\Delta$ and magnetic
field $H$, as shown in Fig.~11(c).
Since we are interested only in the possible symmetry of
Cooper pair, we do not consider the competition among
superconducting state and other ordered phases.
For instance, when the magnetic field becomes large enough,
quadrupole ordering may be dominated,
as deduced from the experimental result.
In the region of $\Delta$$<$0, for zero magnetic field, there appears
the singlet off-site pair state composed of $\Gamma_7$ electrons,
as shown in Fig.~11(a).
However, for $\Delta/t$$<$$-0.8$, the triplet pair state appears only
by applying a small magnetic field.
Since this state is considered to be dominated by the magnetic phase,
superconductivity may disappear in actuality.
In fact, the maximum eigen value in this region is not enhanced
in comparison with that in the region of $\Delta$$\approx$0.
For $\Delta$$>$0, we observe a wide region of singlet state,
which will be dominated the CDW-like phase.
The local $\Gamma_1$ singlet is robust and it is not easily destroyed
by the magnetic field.
Again the maximum eigen value in this region is not enhanced and
superconductivity is not expected to occur, at least,
in the present model without phonons.

Finally, let us focus on the region around $\Delta$$\approx$0.
For $H$=0, there occurs the singlet off-site pair,
composed of a couple of $\Gamma_8$ electrons.
When we apply a small magnetic field in the degenerate region around
$\Delta$$\approx$0, we can observe the odd-parity triplet pair.
The maximum eigen value in this region is clearly enhanced
in comparison with that for $\Delta$$>$0 or $\Delta$$<$0 and thus,
it may be connected to the superconductivity in the thermodynamic limit.
Upon further increasing $H$, the triplet phase eventually disappears
and a singlet-pair state turns to occur, as mentioned in Fig.~11(b).
However, such a state will be dominated by quadrupole ordered phase.

Readers may consider that the change from singlet off-site
to odd-parity triplet pair is related to the second
transition in the superconducting phase of PrOs$_4$Sb$_{12}$
detected by thermal conductivity.\cite{Izawa}
This is an interesting scenario to explain the multiple
superconducting phase, but the second transition
below $T_{\rm c}$ has not been
confirmed by other experimental measurements.
In any case, it seems to be premature to conclude the pairing symmetry
only from the present 2-site calculations.

%
%
\section{Discussion and Summary}

In this paper, we have analyzed two kinds of the Anderson models
with active orbital degrees of freedom by using the NRG technique.
It has been found that magnetic fluctuations significantly remain,
if $\Gamma_4^{(2)}$ triplet is the excited state with small excitation
energy. By further analyzing the Hubbard-like model constructed from 
the $j$-$j$ coupling scheme, we have proposed a scenario that anisotropic
Cooper pair occurs in the interchanged region between singlet and
triplet states.

As mentioned repeatedly in this paper, magnetic fluctuations remain
for the degenerate region in which singlet and triplet ground states
are interchanged.
Such magnetic fluctuations may be antiferromagnetic,
since local triplets easily lead to antiferromagnetic insulating state
in the periodic lattice.
Thus, the off-site singlet pair appearing
in the region of $\Delta$$\approx$0 in Fig.~11(a) may be the $d$-wave
Cooper pair mediated by magnetic fluctuations.
Note that in the exciton mechanism for the superconductivity
of PrOs$_4$Sb$_{12}$, $d$-wave pair was also suggested.\cite{Koga}

However, it is premature to conclude the $d$-wave pair, since
the present calculation has been done only in the 2-site case.
In addition, we are considering the pair in the degenerate region,
where spin {\it and} orbital fluctuations should play some roles.
Note that $\Gamma_8$ electrons have spin and orbital degrees of
freedom.\cite{Hotta1}
In particular, it has been pointed out that triplet pair can be
induced by the cooperation between spin and orbital fluctuations.
\cite{Takimoto2a,Takimoto2}
Within a random phase approximation, it is easy to show that
the spin and orbital fluctuation terms compete each other
for spin-singlet channel, while for spin triplet pairing,
orbital fluctuations are cooperative with spin fluctuations.
Thus, it is considered that the spin-triplet pair occurs
in the orbital degenerate systems.
In fact, by applying a small magnetic field, we have observed
the odd-parity triplet pair, which may be mediated by cooperative
fluctuations of spin and orbital degrees of freedom.
In order to see whether the triplet pair is stabilized or not
for zero magnetic field, it is necessary to solve the Eliashberg
equation in the thermodynamic limit. This is one of future tasks.

We should note that the quadrupole component is also included in
the $\Gamma_4^{(2)}$ triplet, although we have emphasized
magnetic fluctuations in this paper.
In fact, an important role of quadrupole fluctuations has been
suggested experimentally\cite{Kuwahara,Goto} and theoretically.
\cite{Miyake}
In the future discussion on superconductivity,
it is necessary to clarify the different roles of magnetic and
quadrupole fluctuations.

In the present scenario, unconventional superconductivity should
disappear when $\Delta$ becomes large.
For large $\Delta$($>$0), charge degree of freedom is dominant
and it should be coupled to phonons.
If we further include electron-phonon interaction into the
present model, conventional $s$-wave superconductivity may
occur for large $\Delta$.
Namely, when $\Delta$ is increased, unconventional superconducting
state is eventually changed to conventional one.
Such a change has been experimentally observed in
${\rm Pr(Os_{1-x}Ru_{x})_4Sb_{12}}$.\cite{Frederick}
When x is increased from 0 to 1, the excitation energy
between $\Gamma_1$ and $\Gamma_4^{(2)}$ continuously increases
and the superconducting properties are changed from unconventional
to conventional, consistent with our scenario.
It is another future problem to reproduce quantitatively
the change of superconductivity based on the Hubbard model
Eq.~(\ref{HubbardModel}) with electron-phonon interaction.

\section*{Acknowledgement}

The author is grateful to H. Harima, K. Kubo and K. Takegahara
for discussions on the CEF potential of filled skutterudite compounds.
He also thanks W. Higemoto, K. Kaneko, T. D. Matsuda, N. Metoki,
H. Onishi, Y. \=Onuki, Y. Tokunaga, K. Ueda, R. E. Walstedt
and H. Yasuoka for useful discussions on $f$-electron systems.
This work has been supported by Grants-in-Aid for
Encouragement of Young Scientists under the contract No.~14740219
and for Scientific Research in Priority Area ``Skutterudites''
under the contract No.~16037217 from the Ministry of
Education, Culture, Sports, Science, and Technology of Japan.
The author is also supported by a Grant-in-Aid for
Scientific Research (C)(2) under the contract No.~50211496
from Japan Society for the Promotion of Science.
The computation in this work has been done using the facilities
of Japan Atomic Energy Research Institute and
the Supercomputer Center of Institute for Solid State Physics,
University of Tokyo.


\end{document}